\documentclass[prb,preprint]{revtex4-1} 
\usepackage{graphicx}
\usepackage{amsmath}
\usepackage{mathrsfs}
\usepackage{graphicx}
\usepackage{epstopdf}
\usepackage{amsmath}
\usepackage{amssymb}
\usepackage{textcomp}
\usepackage{default}
\usepackage{default}
\newcommand{\be}{\begin{equation}}
\newcommand{\bea}{\begin{eqnarray}}
\newcommand{\bc}{\begin{center}}            
\newcommand{\ee}{\end{equation}}
\newcommand{\eea}{\end{eqnarray}}
\newcommand{\ec}{\end{center}}
\newcommand{\baa}{\begin{eqnarray*}}
\newcommand{\eaa}{\end{eqnarray*}}
\begin{document}
\title{Generalized golden mean and the efficiency of thermal machines}
%\vskip 24pt
\author{}
\affiliation{}
\author{Ramandeep S. Johal}
\email{rsjohal@iisermohali.ac.in}
%johalr69@yahoo.co.in } \\ 
\affiliation{ Department of Physical Sciences, \\
Indian Institute of Science Education and Research Mohali,\\ 
Sector 81, S.A.S. Nagar, Manauli PO 140306, Punjab, India }
%email: rsjohal@iisermohali.ac.in}
\begin{abstract}
We investigate generic heat engines and 
refrigerators operating between two heat reservoirs, for the condition
when their efficiencies are equal to each other. It is shown that
the corresponding value of efficiency is given as the inverse of the 
generalized golden mean, $\phi_p$,
where the parameter $p$ depends on the degrees of irreversibility of both engine 
and refrigerator. The reversible case ($p=1$) yields the
efficiency in terms of the standard golden mean.
We also extend the analysis to a three-heat-resrervoir
setup.
\end{abstract}
\maketitle
\section{Introduction}
Although the golden mean (golden ratio)  
has engaged artists, mathematicians and philosophers since antiquity
\cite{Coxeter, Ogilvy1990, Markowsky1992, Livio}, 
it has been appreciated more recently that it
is not a unique number as far as many of its 
algebraic and geometric properties are concerned
\cite{Falbo2005, Fowler1982}. In fact, one
of the simplest generalizations of the golden mean 
may be defined by the positive solution, $\phi_p$, 
of the following quadratic equation:
\be
x^2 - p x -1 = 0,
\label{x2p}
\ee
where $p$ is a given positive number.
The number $\phi_p$
is given by:
\be
\phi_p = \cfrac{\sqrt{p^2+4} + p}{2},
\label{phip}
\ee
also referred as the $p$th order extreme mean (POEM) \cite{Falbo2005}.
More specifically, when $p$ is a positive
integer $n$, it is addressed as the $n$th order extreme mean (NOEM) \cite{Fowler1982} 
or a member of the family of metallic means.  
For example, $p=1$ gives the golden mean $\phi_1 = (\sqrt{5}+1)/2$; 
$p=2$ yields the silver mean, $\phi_2 = \sqrt{2} +1$, and so on.
%
% Further, $\phi_p$ can be expressed in terms of 
% continued fractions:
% \be
% \phi_p = p + \cfrac{1}{p+ \cfrac{1}{p+ \dotsb}},
% \ee
% or, as nested radicals:
% \be
% \phi_p = \sqrt{1+p \sqrt{1+p \sqrt{1+p}}}\cdots,
% \ee
% which generalize the more well-known representations of 
% the golden mean ($p=1$). 
%
Amongst other mathematical constructions, 
the ratio of successive terms in the generalized Fibonacci
sequence 
$F_{n}^{} = p F_{n-1}^{} + F_{n-2}^{}$,
tend to this number:  
\be
\lim _{n \to \infty} \frac{F_{n+1}}{F_n} = \phi_p.
\ee
It also relates the lengths of diagonals of a regular odd $n$-gon ($n\geq 5$).
 For other properties and identities related to $\phi_p$, see Ref. 
\cite{Fowler1982}.

The generalized golden mean ($\phi_p$)  appears as
the optimal solution determining the shape of Newton's frustum
that faces the least resistance while moving through a rare medium \cite{Sampedro2010}.
The metallic means family has been related to quasiperiodic
dynamics \cite{Spinadel1997}. Another simple physical example
is a semi-infinite resistor network \cite{Srinivasan1992, Kasperski2013} as shown in Fig. 1a,
whose equivalent resistance ($r'$) between points $A$ and $B$ satisfies the quadratic
euquation: $(r')^2-r r' -1 =0$, and therefore, $r' = \phi_r$. 
In this article, we describe an occurrence of the 
generalized golden mean in the context of thermodynamics by 
relating this number to the instance of equal efficiencies of
a heat engine and refrigerator. 
\begin{figure}[ht]
	\includegraphics[width=1.1\linewidth]{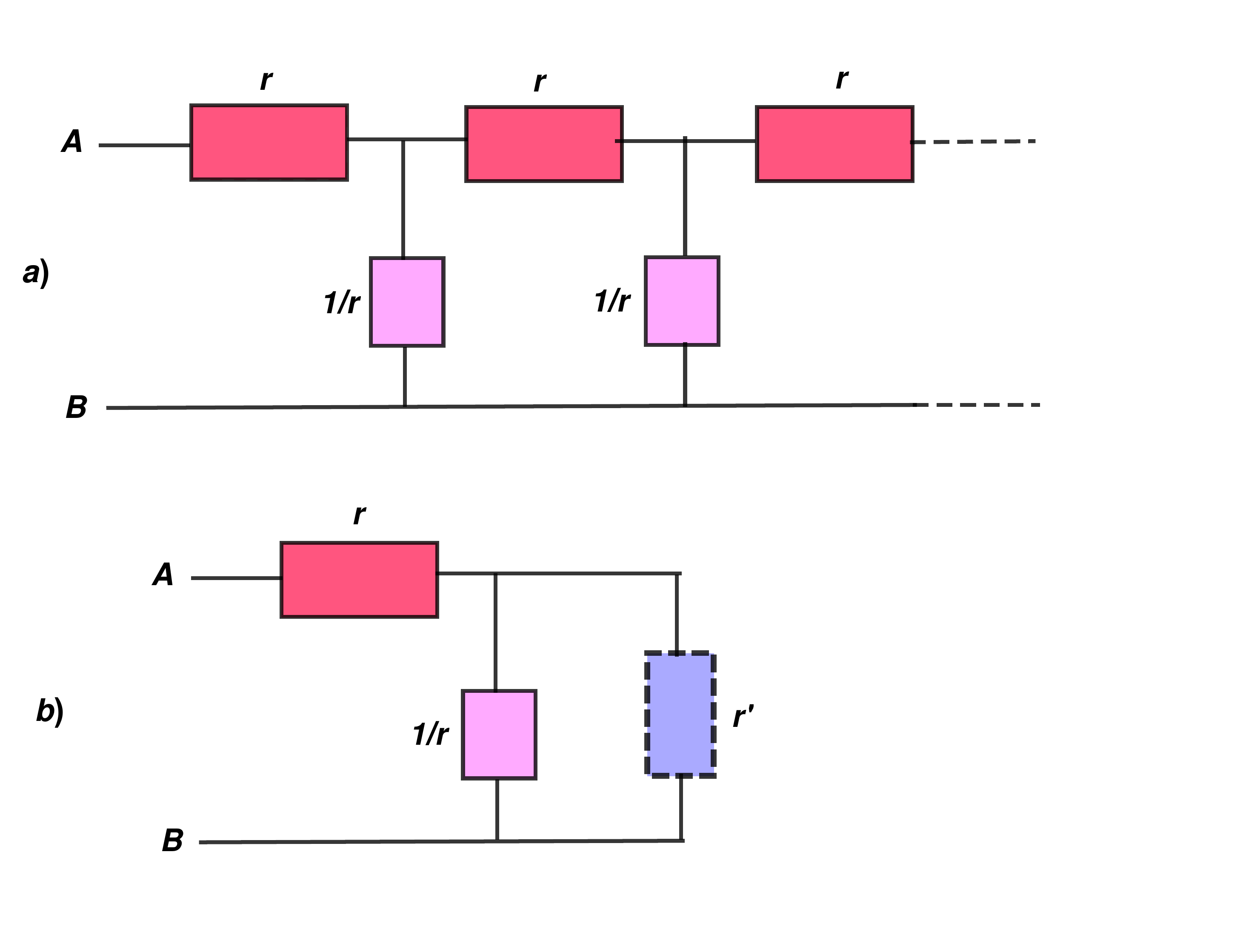}
	\caption{a) An semi-infinite resistor network where each
	resistance in series is $r$ while a resistance in parallel is $1/r$. b)
	The effective resistance between points A and B is $r' = \phi_r$.}
	\label{equalresis}
\end{figure}
\section{Two-heat-reservoir setup}
A heat engine and a refrigerator constitute basic mechanisms
of a thermal machine operating between  
two heat reservoirs at unequal temperatures, say, 
$T_h$ and $T_c$ ($T_c < T_h$) \cite{Zemansky1997}.
An irreversible machine always leads to a net entropy increase
of the universe. On the other hand, the ideal limit of a reversible 
operation implies no net entropy change of the universe. Further, 
irreversibility leads to a decrease in performance of the machine,
and so the efficiency of a thermal machine is limited by its maximal
value, known as the Carnot efficiency, obtained in the reversible case. 

{\it Heat engine}:
We first consider an irreversible heat engine
working in a cycle, and suppose that 
 $W$ amount of work is extracted per cycle when $Q_h$ 
amount of heat is absorbed from the hot reservoir.
The total entropy generated per cycle is:
\be
\Delta_{\rm tot} S = -\frac{Q_h}{T_h} +  \frac{Q_c}{T_c} >0.
\label{dst}
\ee
where $Q_c = Q_h-W$,  
is the heat rejected to the cold reservoir. All 
quantities defined above are positive.
From Eq. (\ref{dst}), we can express
the work output of an irreversible cycle as:
\be
W = (1-\theta) Q_h - T_c \Delta_{\rm tot} S,
\label{wirr}
\ee
where $\theta = T_c/T_h$. 
Let $\Delta_{\rm rev} S = Q_h/T_h$ be the  
entropy transferred  
from the hot reservoir to the working medium, 
in the reversible case. 
Then, the efficiency, $E=W/Q_h$, can be written as 
\be
E  =  1 - \left( 1+ \frac{\Delta_{\rm tot} S}{\Delta_{\rm rev} S} \right) \theta.
\label{efirr}
\ee
Here, we define the ``irreversibility'' parameter, 
$z = 1+ {\Delta_{\rm tot} S}/{\Delta_{\rm rev} S} > 1$, so  
that the efficiency of the engine can be expressed as $E = 1- z \theta$.  
 The reversible case corresponds to 
$\Delta_{\rm tot} S=0$, or $z=1$, yielding
Carnot efficiency equal to $1-\theta$. 
In other words, $0 \le E \le 1-\theta $ implies 
$1\le z \le 1/\theta$.  

{\it Refrigerator}:
Now, let us consider the machine in the refrigerator mode.
Suppose, an input work $\mathscr{W}$ is required to transport
$\mathscr{Q}_c$ amount of heat against the temperature
gradient. Let $\mathscr{Q}_h = \mathscr{Q}_c + \mathscr{W} $, 
be the amount of heat deposited in the hot reservoir.
The efficiency of a refrigerator is defined as $R= \mathscr{Q}_c/\mathscr{W}$.
The total entropy generated per cycle \cite{Zemansky1997} is:
\be
\Delta_{\rm tot} \mathscr{S} = \frac{ \mathscr{Q}_h}{T_h} -  \frac{ \mathscr{Q}_c}{T_c} >0.
\ee
Let $\Delta_{\rm rev} \mathscr{S}$ be the amount of heat  
transferred  reversibly, from the {\it cold} reservoir to the working medium. 
Then we have $\mathscr{Q}_c = T_c \Delta_{\rm rev} \mathscr{S}$, and so,
we can write
\be 
\mathscr{W} = \frac{ 1-\theta}{\theta} \mathscr{Q}_c + T_h \Delta_{\rm tot} \mathscr{S}.
\ee
The efficiency is then given by:
\be 
R = \theta \left(1+ \frac{\Delta_{\rm tot} \mathscr{S}}
{\Delta_{\rm rev} \mathscr{S}} - \theta
\right)^{-1}.
\label{refirr}
\ee
Analogous to the case of engine, we define the parameter 
$z' = 1+ {\Delta_{\rm tot} \mathscr{S}}/{\Delta_{\rm rev} \mathscr{S}} >1$,
and express the efficiency of the refrigerator as : $R = \theta (z' -\theta)^{-1}$.
In the reversible case, $z'=1$, and so $R = \theta (1 -\theta)^{-1}$.
Note that, unlike the parameter $z$, $z'$ is not bounded from above.

In the above, we have identified the expressions for efficiencies
as functions of
the ratio of temperatures as well as the irreversibility parameter
$z$ or $z'$. These expressions 
refer to {\it any} irreversible thermal cycle between 
the two reservoirs, with given values of 
$\theta, z$ and $z'$.  
We may now ask, for what ratio of temperatures,
a heat engine and a refrigerator with the given values
of their respective irreversibility parameters, have 
the {\it same} efficiency, and what is its value?

Thereby, setting $E = R$, and solving for $0<\theta <1$, we
obtain:
\be
\theta = \frac{z z' + 2 -\sqrt{(z z')^2 +4 }}{2z}.
\label{theq}
\ee
Then, the efficiency, $E=1-z\theta$, at the above condition is given by:
\be
E = R = \frac{\sqrt{(z z')^2 + 4} - z z'}{2}.
\label{ezrzp}
\ee
Interestingly, the above expression for the efficiency depends only on the 
{\it product} of the irreversibility parameters. 
Thus, we may define ${z z'} \equiv p$, and 
rewrite Eq. (\ref{ezrzp}) as follows
\be
E = \frac{\sqrt{p^2 + 4} - p}{2} = \frac{1}{\phi_p},
\label{ez}
\ee
where $E$ is expressed in terms of $\phi_p$ from Eq. (\ref{phip}).
Note that, in the above  $p \geqslant 1$, where $p=1$ 
implies the reversible case, with $z= z' = 1$,
yielding $E = R = (\sqrt{5}-1)/2 = 1/\phi_1$, which was earlier noted
 in Ref. \cite{Popkov}. Correspondingly,
the ratio of temperatures in the reversible case is required
to be: $\theta = (3-\sqrt{5})/2 = 2-\phi_1$.

\section{Three-heat-reservoir setup}
Next, we show the occurrence of the generalized golden 
mean in a slightly different setting. 
Let us consider three heat reservoirs 
with temperatures ordered as 
$T_l < T_c < T_h$. Assume that we can operate
an engine between the reservoirs at $T_h$ and $T_l$,
and a refrigerator between the reservoirs at 
$T_c$ and $T_l$ (see Fig. 2). Further, let these
be ideal or reversible machines,
so that the respective thermal efficiencies
are given as:
\be
E_{hl} = \frac{T_h-T_l}{T_h}, \quad R_{cl} = \frac{T_l}{T_c-T_l}.
\ee
\begin{figure}[ht]
	\includegraphics[width=1.2\linewidth]{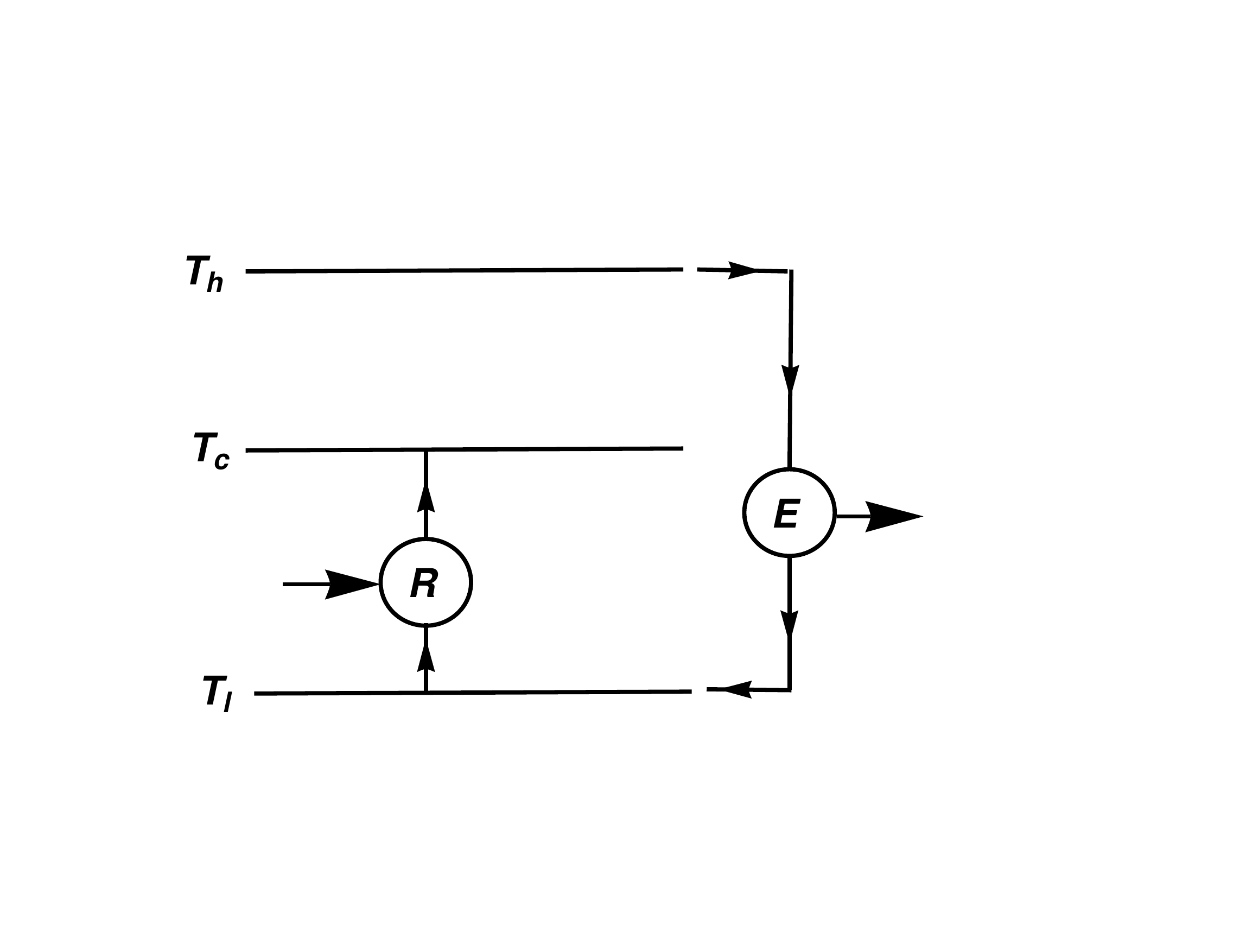}
	\caption{A three heat-reservoir setup, in which 
	a heat engine runs between the hottest ($T_h$) 
	and the coldest ($T_l$)
	reservoirs, while a refrigerator runs between the 
	coldest and an intermediate reservoir ($T_c$). For given 
	 values of $T_h$ and $T_c$, we look for $T_l$ at which 
	 efficiencies of the engine and the refrigerator are
	  equal.}
	\label{3src}
\end{figure}
Now, for given values of $T_h$ and 
$T_c$, we look for the temperature $T_l$ such that 
these two efficiencies become equal.
Setting $E_{hl} = R_{cl}$, we obtain a quadratic
equation for $T_l$:
\be
T_{l}^{2} - (2T_h + T_c) T_l + T_h T_l = 0, 
\ee
whose solutions are:
\be
T_l = \frac{T_h}{2} \left[ \theta +2  \pm \sqrt{\theta^2+4}    \right],
\ee
where $\theta = T_c/T_h$.
Only the negative root above yields an allowed value of the efficiency,   
given by:
\be
E_{hl} = R_{cl} = \frac{ \sqrt{\theta^2 + 4}-  \theta}{2}  
= \frac{1}{\phi_\theta}.
\label{phth}
\ee 
Note that the generalized golden
mean appears above within a reversible setup. Secondly,
due to $0< \theta <1$, it covers a range of parameters
complementary to the expression (\ref{ez}), where 
$p \geqslant 1$. 

Finally, we extend our exploration on the three-heat-reservoir 
case also into the irreversible regime. 
Consider an irreversible engine between $T_h$ and $T_l$,
specified by the irreversibility parameter $z$. Likewise,
let the corresponding parameter for the refrigerator
operating between $T_c$ and $T_l$ be $z'$. Then, following Section II, 
the efficiencies of these machines can be written as:
\be
\bar{E}_{hl} = 1-z\frac{T_l}{T_h}; \quad \bar{R}_{cl} = \frac{T_l/T_c}{z'- T_l/T_c}.
\ee
Now setting the condtion $ \bar{E}_{hl} = \bar{R}_{cl}$, 
we can solve for $T_l$:
\be
T_l = \frac{T_h}{2z} \left[ p\theta + 2 \pm \sqrt{(p \theta)^2 + 4}
\right],
\ee
where $p = z z'$.
Consequently, the allowed solution for efficiency is given by:
\be
\bar{E}_{hl} =  
\frac{\sqrt{(p \theta)^2+ 4}-p \theta}{2} = 
\frac{1}{\phi_{p\theta}^{}}.
\ee
For $p=1$, we obtain the reversible case discussed above (Eq. (\ref{phth})). Moreover,
letting $\theta \to 1$, which may be realized by taking $T_c \to T_h$, we revert to the case of two reservoirs 
(at $T_h$ and $T_l$), as discussed in Section II.

\section{Conclusions}
In the above, we have observed that the generalized
golden mean determines the efficiencies 
of a heat engine and a refrigerator 
when the latter are set equal to each other. 
When both the machines are reversible, the 
efficiency is related to the standard golden
mean. To extend to irreversible domain,
we have first recast the efficiency of a generic
 heat engine and a refrigerator in
terms of an irreversibility parameter and the
ratio of the reservoir temperatures. 
Then, the efficiency depends
only on the product of the two irreversibility parameters. 
The importance of this step can be appreciated by
noting that {\it only} for the reversible case, we have
$R = Q_c/W_h$ along with $E = W/Q_h$, where $Q_c = Q_h -W$,
i.e. the same amounts of heat and work appear for a {\it reversible} heat
engine as well as for a refrigerator. So, in this case, 
we can express the condition $E=R$, as follows:
\be
\frac{W}{Q_h} = \frac{Q_c}{W}.
\ee
From this, we obtain the equation: $E = E^{-1} - 1$,
whose solution is $E = R = 1/\phi_1$ \cite{Popkov}. 
However, the efficiencies in the irreversible case are $E= W/Q_h$
and $R = \mathscr{Q}_c/\mathscr{W}_h$,
which don't seem useful for applying the $E=R$ condition.
Thereby, the forms (\ref{efirr}) and  (\ref{refirr})
have been employed. Further, we have also extended this
condition to three-reservoir scenario which 
also includes the limiting case of two-reservoir
setup. It will be interesting to identify physical 
situations leading to the equality of
the efficiencies of engine and refrigerator
under the given conditions.
Here, we have discussed one possible setup using three heat reservoirs. 
The interested reader can identify other 
engine-refrigerator pairs in the 
three-reservoir setup which lead to equal efficiencies,
expressed in terms of suitable generalized golden means.
\end{document}